# Trapped air metamaterials for ultrasonic sub-wavelength imaging in water

Stefano Laureti[1*], David A. Hutchins[2], Lorenzo Astolfi[2], Richard L. Watson[2], Peter J. Thomas[2], Pietro Burrascano[3], Luzhen Nie[4], Steven Freear[4], Meisam Askari[5], Adam T. Clare[6] and Marco Ricci[1]

[1] Department of Informatics, Modeling, Electronics and Systems Engineering, University of Calabria, Via Pietro Bucci, 87036, Arcavacata di Rende (CS), Italy.
{stefano.laureti, marco.ricci}@unical.it

[2] School of Engineering, University of Warwick, Coventry, CV4 7AL, UK.
{D.A.Hutchins, lorenzo.astolfi, r.watson.2, P.J.Thomas}@warwick.ac.uk

[3] Department of Engineering, University of Perugia, Polo Scientifico Didattico di Terni, Via di Pentima 4, 05100, Terni, Italy. pietro.burrascano@unipg.it

[4] School of Electronic and Electrical Engineering, University of Leeds, Leeds, LS2 9JT, UK.
{L.Nie, s.freear}@leeds.ac.uk

[5] School of Physics and Astronomy, University of St Andrews, St Andrews, KY16 9SS, UK.
Ma297@st-andrews.ac.uk.

[6] Department of Mechanical, Material and Manufacturing Engineering, University of Nottingham, University Park, Nottingham, NG7 2RD, UK.
Adam.Clare@nottingham.ac.uk

[*]Correspondence to Stefano Laureti (stefano.laureti@unical.it).

**Acoustic metamaterials[1-3] constructed from conventional base materials can exhibit exotic phenomena such as negative refractive index[4], extraordinary transmission/absorption[5,6] and sub-wavelength imaging[7-9]. These are typically achieved by combining geometrical and resonance effects. Holey-structured acoustic metamaterials have already shown potential for sub-wavelength acoustic imaging in air[10,11], where it is relatively simple to achieve the required difference in acoustic impedance needed to create resonances within the holes. However, the use of polymers for metamaterial operation in water is difficult, due to the much lower difference in acoustic impedance between water and the many polymers used, for example, in additive manufacturing. Hence, metals are commonly used[12]. Here we show that this can be**

**overcome by trapping air layers so that they surround each water-filled channel, making sub-wavelength imaging in water possible. The operation of such a "trapped air" design is confirmed at 200-300 kHz ultrasonic frequencies via both finite element modelling and experimental measurements in a water tank. It is also shown quantitatively that the trapped air design outperforms its metal counterpart. The results indicate a way forward for exploiting additive-manufacturing for realising acoustic metamaterials in water at ultrasonic frequencies.**

**Introduction**

In nature, electromagnetic and acoustic macroscopic properties of any materials are governed by their atomic and molecular structures. Those properties are typically characterized by a number of parameters such as electrical permittivity $\epsilon$ and magnetic permeability $\mu$ for electromagnetic waves, or by the density $\rho$ and bulk modulus $K$ for acoustics. These parameters normally assume positive values for natural materials. In 1968, Veselago[13] showed theoretically that phenomena such as negative refractive index could be obtained when these parameters are negative. Acoustic metamaterials (AMs) can exhibit exotic properties by using sub-wavelength features (referred as meta-atoms) in the form of resonators and scatterers. These are typically arranged in a periodic way so that they behave like a bulk continuous material but with 'on-demand' effective properties[1,14,15]. For example, Zhang et al.[2] and Yang et al.[3] describe materials exhibiting negative effective density $\rho_{eff}$ and bulk modulus $K_{eff}$, resulting in a negative acoustic refractive index ($\eta_{eff} < 0$).

The manipulation of $\rho_{eff}$ and $K_{eff}$ for a given frequency ($f$) and wavelength ($\lambda$) is of interest for ultrasonic imaging in applications such as nondestructive evaluation[16-19] and diagnostic biomedical imaging[20-22]. This arises because of the so-called "diffraction limit", a consequence of the inability to capture the evanescent field which carries the finer sub-wavelength details of an image. The minimum feature scale $w$ that can be distinguished during an ultrasonic test is given via the relation[13]

$$w \cong c/(2 \cdot f) \cong \lambda/2, \qquad (1)$$

where $c$ is the speed of sound of the medium. AMs can be used to overcome this diffraction limit, because the resolution of the image will be dictated by their sub-wavelength internal structure and the resulting exotic effective properties. Holey-structured acoustic metamaterials (HSAMs) produce this enhanced sub-wavelength imaging resolution by the coupling of evanescent waves via Fabry-Pérot Resonance (FPR) mechanisms[9-11]. One consequence for biomedical imaging is that lower frequencies could be used with an AM to obtain the same imaging resolution as that from a conventional measurement at higher frequencies, and this would allow an increased penetration into the body.

In their pioneering work[25], Christensen *et al.* showed that FPRs exist within the holes of a 2D array of square-shaped apertures fabricated within a bulk material. Consider a HSAM having a thickness $h$, through-thickness channels of width $a$, and a distance between hole centres, *i.e.* the lattice constant, $\Lambda > a$. Usually $h > \Lambda > a$. The transmission process of acoustic waves through such HSAM is regulated by the resonant mode within each hole, and the transmission coefficient $T^{00}$ is unity when resonances occur within the through-thickness channels so that

$$\lambda_{FPR_m} = \frac{2 \cdot h}{m} \rightarrow |T^{00}| = 1, \qquad (2)$$

where $\lambda_{FPR_m}$ is the wavelength at resonance and $m = 1, 2, \ldots, N$ with $N \in \mathbb{N}$.

Equation (2) means that all the acoustic information, including that from evanescent waves, is transferred from one side of the metamaterial to the other whenever the perpendicular component of the incident field wavevector $k_\perp$ is such that $\frac{2\pi}{k_\perp} = \lambda_{FPR_m}$, as the transmission coefficient has a modulus of unity[10]. Under this hypothesis, the resolution is not dictated by the global material properties but by the lattice constant $\Lambda$. In fact, by considering the AM a periodic structure along the surface plane, $\pi/\Lambda$ determines the limit of the Brillouin zone, so the smaller is $\Lambda$, the larger is the Brillouin zone and $\Lambda/a$ determines also the $\rho_{eff}$ of the HSAM[10-11,26-30]. Examples of imaging in air at resolutions of up to $w/50$ at audible frequencies have been reported[10,11]. However, if HSAMs are to be used for acoustic sub-wavelength imaging in water, it is essential that there is a sufficient difference in acoustic impedance $Z$ between the bulk material ($Z_b$) of the AM and the water-filled holes ($Z_{water}$) within which the acoustic signal is contained. Metallic substrates could be used; however, there is an attractiveness to using polymers as they are readily translated to additive manufacturing technologies, where the use of metallic substrates is limited due to the high aspect ratio $h/a$ of the holes and their relatively-high cost.

Although some work has been reported showing polymer-based HSAMs operating in water[23,24], the effects of acoustic coupling from the water-filled holes into the polymer substrate would be expected to degrade their performance (see Supplementary Information). As an example, Amireddy *et al.*[24] demonstrated operation in water at $f = 250$ kHz, but significant unwanted acoustic energy within the solid was seen, and an amplitude enhancement due to the additional evanescent wave contribution to the image was not observed, as would be expected for efficient operation of an AM (see Supplementary Material)[10]. Note that Estrada *et al.*[12] reported that polymer substrates would not be expected to be efficient for producing such devices.

In this work, we overcome the above-mentioned limitations and demonstrate that polymer-based HSAM structures can in fact be constructed for efficient use in water, using a completely new "trapped air" concept. As the acoustic impedance of air ($Z_{air}$) is very low (so that $Z_{air} \ll Z_b, Z_{water}$), acoustic isolation between air and both water and a typical polymer is very high, helping to address the cross-coupling problem. The properties of this design are demonstrated here using both Finite Element Model simulations and experimental measurements. A thorough comparison of the proposed design with its conventional polymeric and metallic counterparts is reported in the Supplementary Information, where the present design is further shown to be more effective.

**Trapped Air Holey-Structured Metamaterial Design**

The new design, hereinafter referred as Trapped Air HSAM was developed to allow a layer of air to completely surround each water-filled channel, in a way such that metamaterials could be constructed in polymer using standard additive manufacturing technologies. Full details are given in the Supplementary Information, but Figure 1(a) shows the construction of the two layers. This was designed with $\Lambda$ = 1.2 mm, $a = 0.8$ mm, and $h = 6$ mm, producing FPRs in the $f = 100 - 300$ kHz range so as to match the available ultrasonic transducer operating bandwidth (see Methods). The bottom layer to the left contains an array of hollow polymer rods, in this case with an air spacing between them. The second layer, shown to the right of Figure 1(a), contains an array of holes to match the channels in each of the hollow rods. This is inverted and placed on top of the first layer and the seal between the two layers made watertight. When immersed in water, the channels would then fill with water, but each would be separated from the other by the air trapped within the sealed device.

A schematic diagram of an assembled Trapped Air HSAM is shown in Figure 1(b). It can be seen how the air layers act to insulate the water-filled channels acoustically from each other.

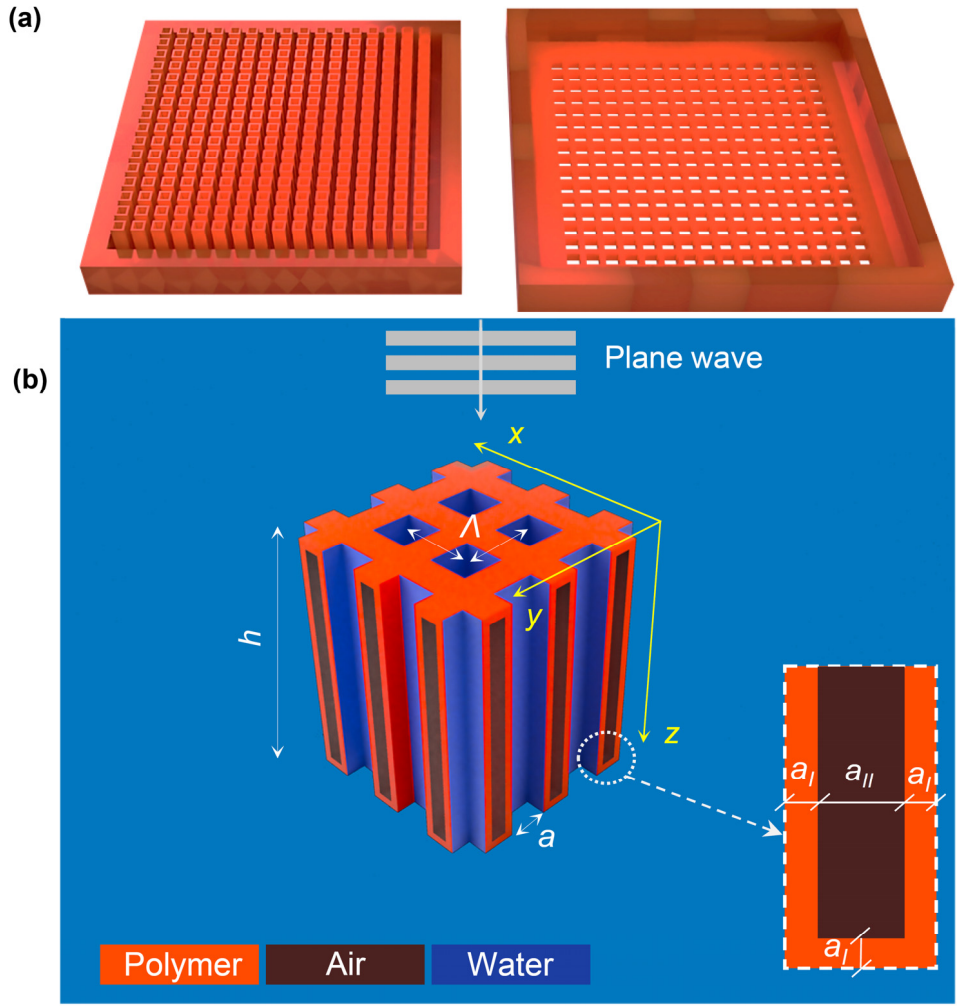

Figure 1: (a) The two additively-manufactured layers which, when placed together, form the Trapped Air HSAM; (b) a sketch representing four unit cells of holes of the Trapped Air HSAM. Here $\Lambda$ = 1.2 mm, $a$ = 0.8 mm, and $h$ = 6 mm. Air was trapped within the polymer substrate as shown – a zoomed diagram of the embedded air cavity is depicted to the right, where $a_I$ = 0.1 mm and $a_{II}$ = 0.2 mm.

A Trapped Air HSAM was constructed via additive manufacturing. The unit cell dimensions were as in Fig. 1(b) but containing an array of 16 x 16 holes. The device was placed in water at room temperature together with an E-shaped aperture machined into a ~0.9 mm thick brass plate. The arms of the "E" aperture had a width of 1 mm. At the expected simulated FPR of 246 kHz, λ in water is ≈ 6 mm. Hence, for operation at or close to this frequency the aperture was of λ/6 in size, which is beyond the conventional diffraction limit. An ultrasonic piezoelectric transducer was placed at ~150 mm from the brass slab and fed with a frequency modulated "chirp" signal over the 100-400 kHz frequency range. A miniature hydrophone (Precision Acoustics) of 0.2 mm active diameter collected the ultrasonic pressure field transmitted through the metamaterial at various locations across the *x-y* plane, see Fig.2.

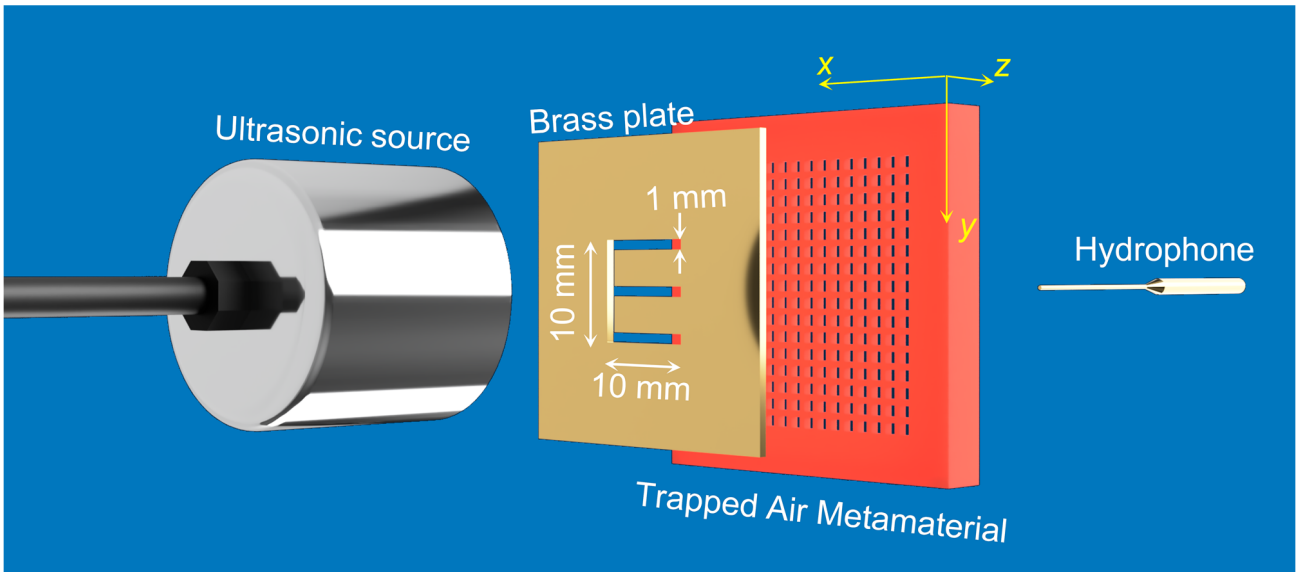

Figure 2: Experimental setup, showing the size and orientation of the "E" aperture, as well as the location of the hydrophone and ultrasonic source.

The received signals were processed with Chirp Z-Transform algorithm[31] and images obtained for comparison to FEM predictions.

**RESULTS**

**Finite Element Simulations**

A 2-D FEM was built to obtain the transmission coefficient $|T^{00}|$ of two HSAM designs – one containing trapped air, and one with simply holes within a solid substrate (each having the same values of $\Lambda$, $a$ and $h$ as shown in Fig. 1(b)). A plane wave acoustic field propagating along the *z*-axis direction was employed in the modelling together with a perfectly- matched layer domain. Predictions for $|T^{00}|$ were obtained over the $f = 100 - 300$ kHz range and are plotted in Fig. 3. It is clear that new trapped air design lead to the successful establishment of FPRs, whilst a flattened behaviour is obtained for a standard polymer design. This indicates that, without this acoustic isolation, a severe disruption to the operation of the metamaterial is likely to occur. The reader is referred to Supplementary Information for details about the FEM methodology and comparison to a conventional HSAM with a metallic base.

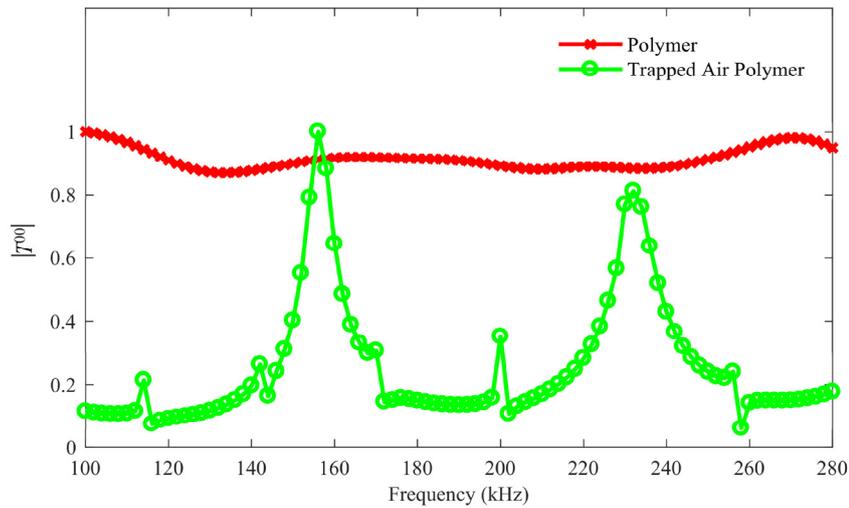

Figure 3: Transmission coefficient $|T^{00}|$ for designs with and without trapped air obtained via 2-D FEM simulations.

**Subwavelength imaging**

This is demonstrated experimentally in Fig. 4(a), whereby the subwavelength features of the E-shaped aperture (1 mm) are faithfully reconstructed at the resonance frequency using the Trapped Air HSAM. The experimental findings are further corroborated by a 3-D FE simulation in which the subwavelength imaging experiment was simulated, see Fig. 4(b). Finally, Fig. 4(c) shows a cross-section of the amplitude obtained at $x = 10$ mm for both the experiment and the simulations. A good agreement between the two is found, as the "E" profile is well-reconstructed. The reader is referred to Supplementary Information for further details concerning the 3-D model and experiment, a comparison with a metal HSAM counterpart, and additional experimental and modelling results which illustrate the unsuitability of a standard polymer HSAM for use in water. This further demonstrates the capability and potential of the Trapped Air design strategy.

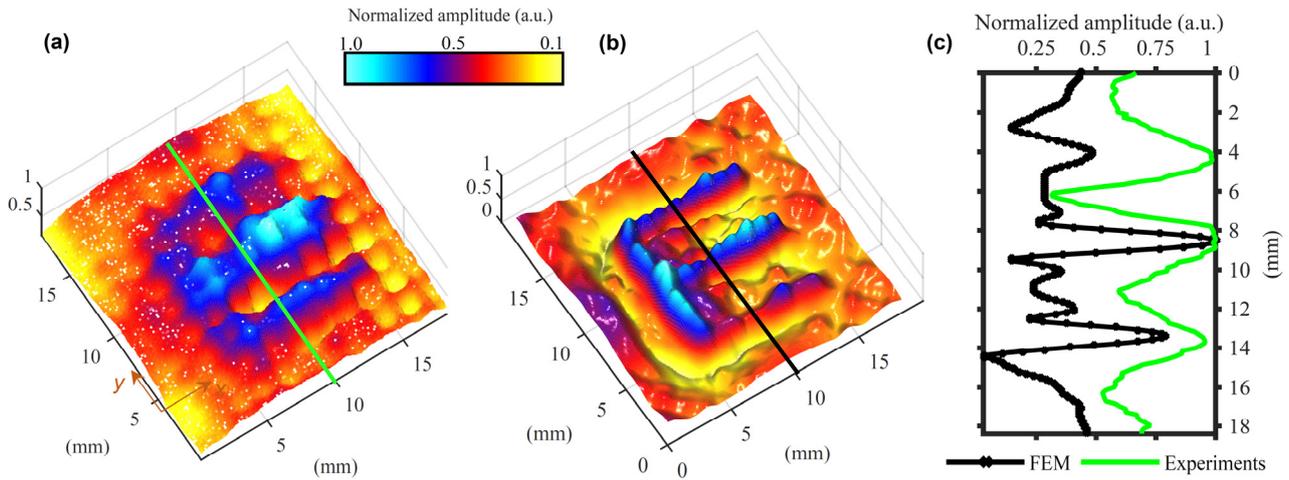

Figure 4: Comparison of (a) experimental results in a water tank and (b) FEM simulations for the Trapped Air HSAM; (c) a comparison plot of the normalized pressure amplitude at $x$ =10 mm, corresponding to the green and black lines in (a) and (b) respectively.

## CONCLUSIONS

It has been shown that the use of trapped air within an acoustic metamaterial can enable such structures to be fabricated from polymers in a standard additive manufacturing process. Their performance has been modelled and shown to be similar to that obtained experimentally at ultrasonic frequencies in water. As expected, the air layers have prevented acoustic cross-coupling from the water-filled channels into the polymer substrate, thus enabling a new class of acoustic metamaterial to be considered for ultrasonic sub-wavelength imaging, sound shielding/extraordinary transmission, and in other acoustic metamaterial designs where high-impedance mismatches between channels are strongly desired[2-3,32].

# METHODS

A detailed description of the materials and methods employed in the present research is reported below.

**2-D Finite Element Modelling simulations**

A 2-D Pressure Acoustics Frequency Domain Finite Element Model was realised in COMSOL Multiphysics® to characterise the frequency behaviour of the tested HSAMs. Two different geometric arrangements were tested: one for simulating either polymer or Nickel HSAMs, while the latter employed a volume of trapped air enclosed in a polymer shell. The geometry employed for the polymer and nickel cases is illustrated Fig. 5.

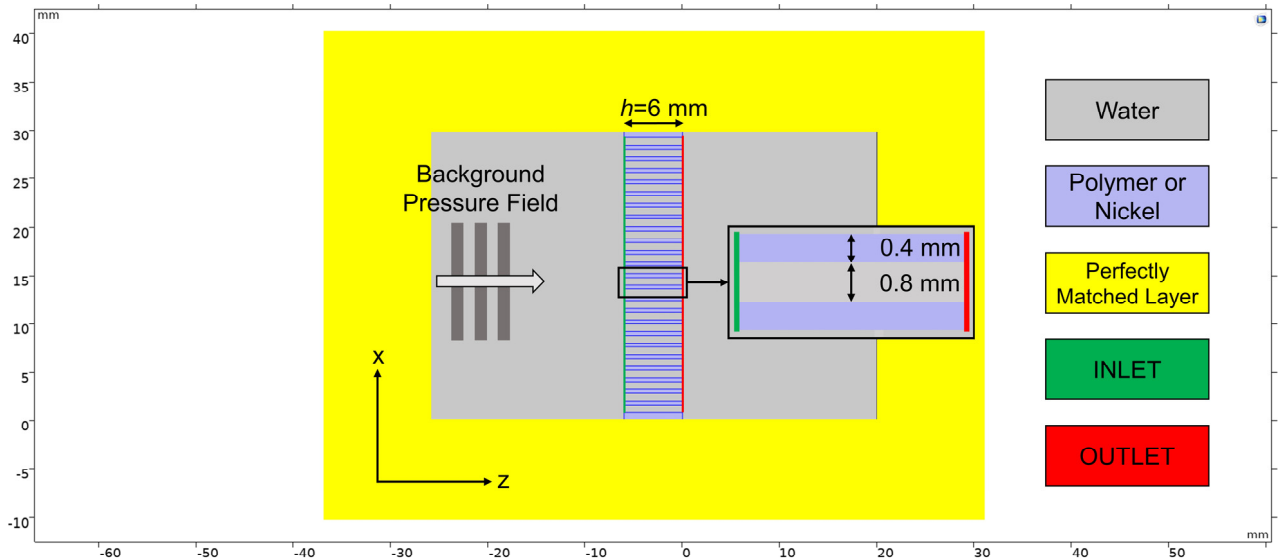

Figure 5: 2-D FEM simulation geometry of a polymer/nickel HSAM with 24 holes.

A control volume of 20 mm was used both before and after the metamaterial in the *z*-direction, and the whole geometry was surrounded by a 10 mm Perfectly Matched Layer (PML) to simulate an infinite water tank. The geometry was meshed so to have at least 10 elements to represent each channel inlet and outlet. A 1 Pa (peak-to-peak) plane wave travelling along the *z* direction was incident onto the AM at different frequencies in the 100-300 kHz range (at 100 Hz increments). The Transmission coefficient $|T^{00}|$ of each structure was obtained as the ratio of the outlet pressure energy to that at the inlet. A different geometry was employed to demonstrate the effect of the trapped air inside the polymer, represented in white in Fig.6.

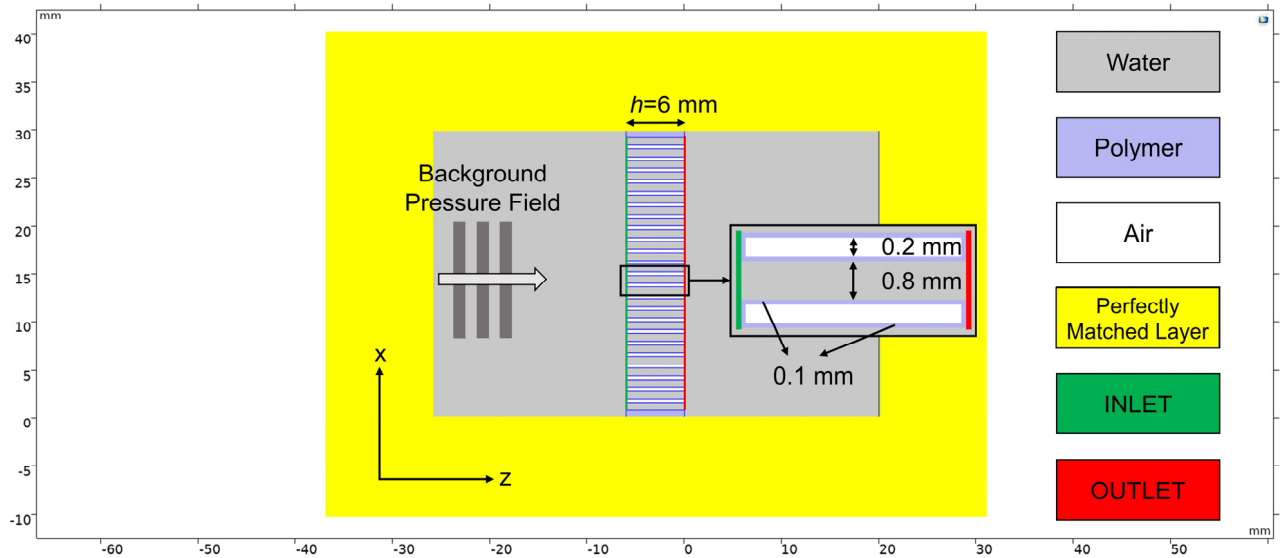

Figure 6: 2-D FEM simulation geometry Trapped Air Polymer containing 24 holes.

**3D Finite Element Modeling simulations for Sub-wavelength imaging**

The 3D model consisted of a plane wave travelling through an "E" shaped aperture having sub-wavelength thickness, carved out of a 1 mm thick brass plate. A plane wave radiation condition was applied to the first boundary in *z* direction and a soft boundary condition was applied to the last boundary in *z* direction to avoid reflections. Data were analysed by imaging the pressure field amplitude at 0.1 mm from the HSAMs outlet, see Figure 7.

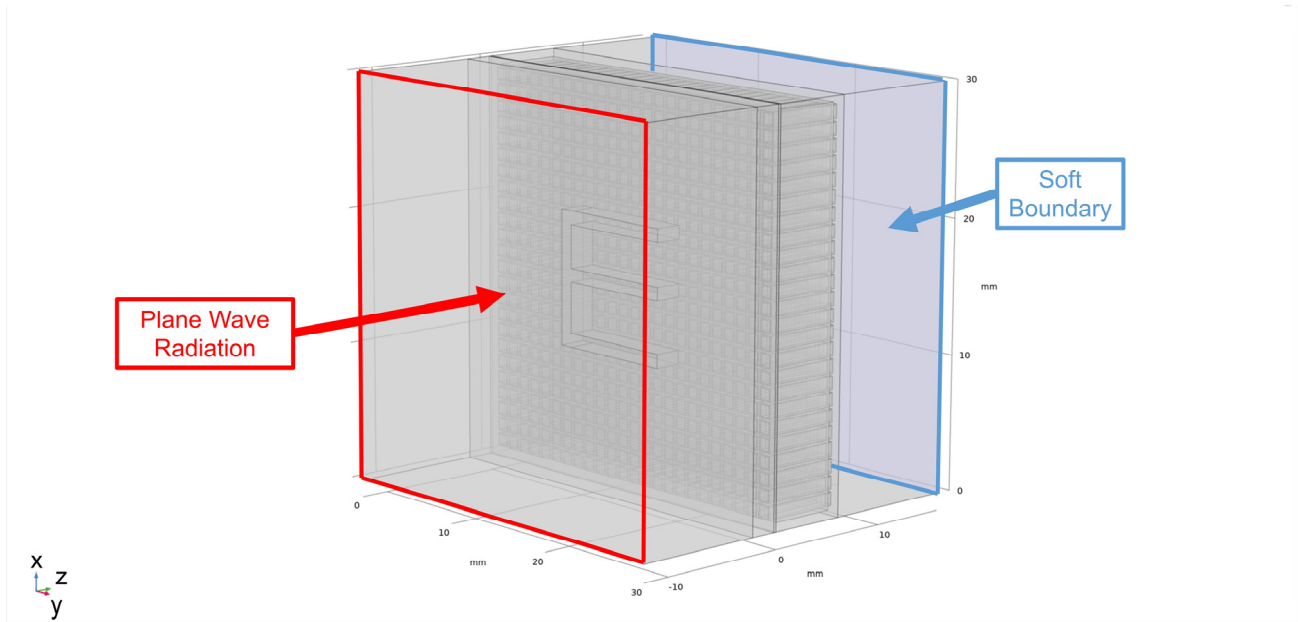

Figure 7: 3-D FEM simulations geometry for the study of sub-wavelength imaging.

**HSAMs Additive Manufacturing**

The Trapped Air HSAM was successfully manufactured containing a 16×16 array of holes with $\Lambda$ = 1.8 mm and $a_I = 0.2$ mm (slightly different from the target values due to manufacturing tolerances). This used a Daylight precision hard white resin and a Liquid Crystal Precision 1.5 3-D printer from Photocentric Group. A HSAM sample without trapped air was also printed in the same material, and contained 24×24 water-filled channels. As an additional comparison, a nickel alloy sample of the same geometry was designed in a CAD software environment and sliced using Materialize Magics 2.1 before fabrication with a Renishaw AM250 SLM Powder Bed Fusion printer. This had a laser power of 200 W and a laser spot size of 150 μm.

**Experimental Setup**

The experiments were conducted in a 400 mm long, 350 mm wide and 200 mm deep volume of water enclosed in a custom-made acrylic tank. The input voltage signal was generated by a NI PXI-5421 Arbitrary Waveform Generator, amplified by a NCA1000-2E amplifier and emitted by a custom made 25.4 mm diameter piezocomposite transducer. An ultrasonic chirp signal of 50 μs duration and with frequencies sweeping from 100 to 400 kHz was used. The radiated ultrasonic signal passed through a 1 mm thick plate containing the subwavelength aperture in the form of the letter "E". The through-transmitted energy was then collected by the 6 mm thick Trapped Air HSAM, and transferred to its far side to be then acquired at a distance of ~0.15 mm from the outlet surface using a 0.2 mm diameter Precision Acoustics needle hydrophone. Signals were then captured using a National Instruments PXI-5122 14 Bit Digitizer sampling at a rate of 100 MS·s$^{-1}$. Both the Waveform Generator and the Digitizer were enclosed in a NI PXI-1042Q chassis. A grid of 131×131 measurement points was acquired by means of a 3D motorized stage with steps of 0.14 mm for an area of 18.34×18.34 mm$^2$. In order to approximate a flat far field wave front, the distance between source and hydrophone was chosen to be 150 mm. Note that the setup employed is similar to that described in[33].

For each scanned point, the DC component was removed from the time domain waveform and the region of interest, selected via a rectangular time window, was converted into frequency domain data via a Chirp Z-Transform algorithm[31].

**Data availability**

The datasets generated during and/or analysed during the current study are available from the corresponding author on reasonable request.

**Code availability**

The FEM simulations were performed using COMSOL 5.4 (licence no. 7077568). The code is available from (lorenzo.astolfi@warwick.ac.uk) on reasonable request.

**Acknowledgements**

The Authors are thankful to Mr. Frank Courtney (Senior Technician, School of Engineering, University of Warwick, UK) for his invaluable help in the 3D printing of the polymer and trapped-air metamaterial. Funding for this work was provided through the UK Engineering and Physical Sciences Research Council (EPSRC), Grant numbers EP/N034163/1, EP/N034201/1 and EP/N034813/1.

**Authors Contributions**

Dr. Stefano Laureti conceived the idea of trapping air inside a Holey-Structured Acoustic Metamaterial, designed and performed experiments and finite element modelling, interpreted and compared the experimental data and the model, wrote and revised the manuscript and guided the project. Prof. David A. Hutchins and Prof. Marco Ricci supervised, guided, founded the project and revised the paper. Prof. Marco Ricci has also developed the post-processing scripts for experimental data interpretation. PhD student Mr. Lorenzo Astolfi designed the metamaterials, performed experiments and finite element modelling, wrote and revised the paper. Dr. Richard L. Watson supervised the design and fabrication of the metamaterials and revised the manuscript. Prof. Peter J. Thomas gave conceptual advice and revised the paper. Prof. Pietro Burrascano reviewed the work. Dr. Luzhen Nie, Prof. Steven Freaar, Dr. Meisam Askari and Prof. Adam Clare were project collaborators, the last two also being in charge of the design and fabrication of the metal sample.

**Corresponding author**

Correspondence to Stefano Laureti (stefano.laureti@unical.it).

**Additional information**

The authors declare no competing financial interests and no competing interests. Supplementary Information accompanies this paper. Reprints and permissions information is available online at http://npg.nature.com/reprintsandpermissions.

# Supplementary Information

**Holey-Structured Acoustic Metamaterial: Theoretical Background and Acoustic impedance mismatch**

Christensen[1] *et al.* showed that Fabry–Pérot Resonances (FPRs) existed within the holes of a 2D array of square-shaped apertures realized in a bulk material. It was found that the peak frequency $f_{FPR_m}$ of the FPRs was related to the thickness $h$ of the metamaterial itself. A thorough theoretical background on Holey-Structured Acoustic Metamaterials (HSAM) can be found in Zhu *et al.*[2] and Laureti *et al.*[3]. The mathematical details are summarized below to better understand the problem of using polymer to construct HSAM for use in water and to highlight the advantages of the Trapped Air HSAM concept.

The main findings are that under the hypothesis that the wavelength of the incident plane wave $\lambda$ is much larger than the channel's lattice constants $a$ and $\Lambda$, the propagation within the holes via FPRs regulates the acoustic energy transmission process. Thus, for wavelengths corresponding to FPRs, *i.e.* $\frac{2\pi}{k_\perp} = \lambda_{FPR_m}$ with $m = 1, 2, \ldots, N$.   $N \in \mathbb{N}$, an incident plane wave of parallel momentum $\vec{k}_\parallel = (k_x, k_y)$ can be fully-transmitted to the HSAM's outlet. This is shown in equation (1), whereby it is pointed out that the transmission coefficient $T^{00}$ reaches the unitary value when FPRs occur:

$$T^{00}(\lambda_{FPR_m}, \vec{k}_\parallel) = (-1)^m, \qquad (1)$$

which is valid for all values of $\vec{k}_\parallel$. Hence, even those values associated with evanescent waves are successfully transferred to the other side of the metamaterial under FPRs conditions. This is the key point in understanding the working principle of holey-structured acoustic metamaterials: they allow the near field region of the acoustic field scattered by an object, *i.e.* the "E" of the present paper, to be transferred over distances at which it would not normally be possible to be observed.

It should be noted that equation (1) is a consequence of the assumption that the substrate material of the HSAMs is a hard solid, meaning that an infinite acoustic impedance difference exists between the medium inside the hole and the substrate. Under such a condition, the acoustic energy would be totally confined within the channels, so as that each would resonate as a stand-alone cavity when excited by an acoustic field matching the FPR mode. However, conventional materials such as polymers or metals have a finite value of impedance, resulting in an exchange of acoustic energy between the cavities. Note that some sub-wavelength imaging capabilities are likely to be lost if this happens, as the effect relies on independence of the channels to preserve the conditions for evanescent wave interaction.

To gain a quantitative insight into the impedance mismatch problem, consider both nickel ($\rho_{nickel} = 8,900 \text{ kg} \cdot \text{m}^{-3}$, $c_{nickel} = 5,600 \text{ m} \cdot \text{s}^{-1}$) and a typical polymer ($\rho_{polymer} = 1,400 \text{ kg} \cdot \text{m}^{-3}$, $c_{polymer} = 1,500 \text{ m} \cdot \text{s}^{-1}$), which are effectively the employed base materials in this research. Their acoustic impedance values are:

$$Z_{nickel} = \rho_{nickel} \cdot c_{nickel} = 49.8 \times 10^6 \text{ kg} \cdot \text{m}^{-2}\text{s}^{-1},$$

$$Z_{polymer} = \rho_{polymer} \cdot c_{polymer} = 2.1 \times 10^6 \text{ kg} \cdot \text{m}^{-2}\text{s}^{-1}.$$

(2)

When used in air at room temperature ($\rho_{air} = 1.2 \text{ kg} \cdot \text{m}^{-3}$, $c_{air} = 343 \text{ m} \cdot \text{s}^{-1}$), the acoustic impedance mismatch of the two materials with respect to that of the air $Z_{air}$ is:

$$\frac{Z_{nickel}}{Z_{air}} = 1.2 \times 10^5;$$

$$\frac{Z_{polymer}}{Z_{air}} = 5.1 \times 10^3.$$

(3)

Equation (3) shows that both polymer and nickel have a large acoustic impedance mismatch to air, resulting in efficient confinement of acoustic energy within each single channel in both cases[2-6]. However, if the same polymer was used with water-filled channels, where the values for water are assumed to be $\rho_{water} = 1000 \text{ kg} \cdot \text{m}^{-3}$ and $c_{water} = 1480 \text{ m} \cdot \text{s}^{-1}$, we obtain

$$\frac{Z_{polymer}}{Z_{water}} = 1.4, \quad (4)$$

so that now energy will pass much more easily from water into the polymer substrate. However, if a layer of air is trapped between the channels and the polymer substrate – as introduced in this work - the impedance mismatch between air and water is approximately

$$\frac{Z_{water}}{Z_{air}} = 3.6 \times 10^3, \quad (5)$$

*i.e.* of the same order of magnitude as in equation (3), meaning that the novel strategy can be used successfully in water. This can be investigated in more detail via Finite Element Modelling, as showed in the next section. Note that the Trapped Air strategy is expected to outperform the use of metal (nickel) for constructing the metamaterial; this is because:

$$\frac{Z_{nickel}}{Z_{water}} = 33.6 \ll \frac{Z_{water}}{Z_{air}}. \quad (6)$$

For a last overview of the impedance mismatch issue, the theoretical transmission coefficients values $t_{1-2} = 1 - [(Z_2 - Z_1)/(Z_2 + Z_1)]^2$ for the above mentioned combinations are:

$$t_{polymer-water} = 0.97;$$

$$t_{nickel-water} = 0.11; \quad (7)$$

$$t_{air-water} = 0.0011.$$

The very low value of $t_{air-water}$ is thus likely to improve the performance when compared to the use of simple metal or polymer substrate.

**The impedance mismatch problem: 2-D Finite Element Modelling**

A series of 2-D FEM simulations have been conducted to both obtain $T^{00}$ and to gain more insight into the impedance mismatch problem for polymer, nickel and Trapped Air HSAMs (see the Methods section for more details of the FEM model). Two different FEM model geometries have been used to predict the behaviour of polymer and nickel HSAMs and the current trapped-air design - see Fig. 5 and Fig. 6 respectively in the Method section.

The predicted magnitude of $T^{00}$ for water-filled channels is shown in Fig.1 for each case. It is found that FPRs are well-established for both nickel and Trapped Air HSAMs. Lower amplitude resonance peaks are found for the trapper air layers, but these are again at the expected frequencies.

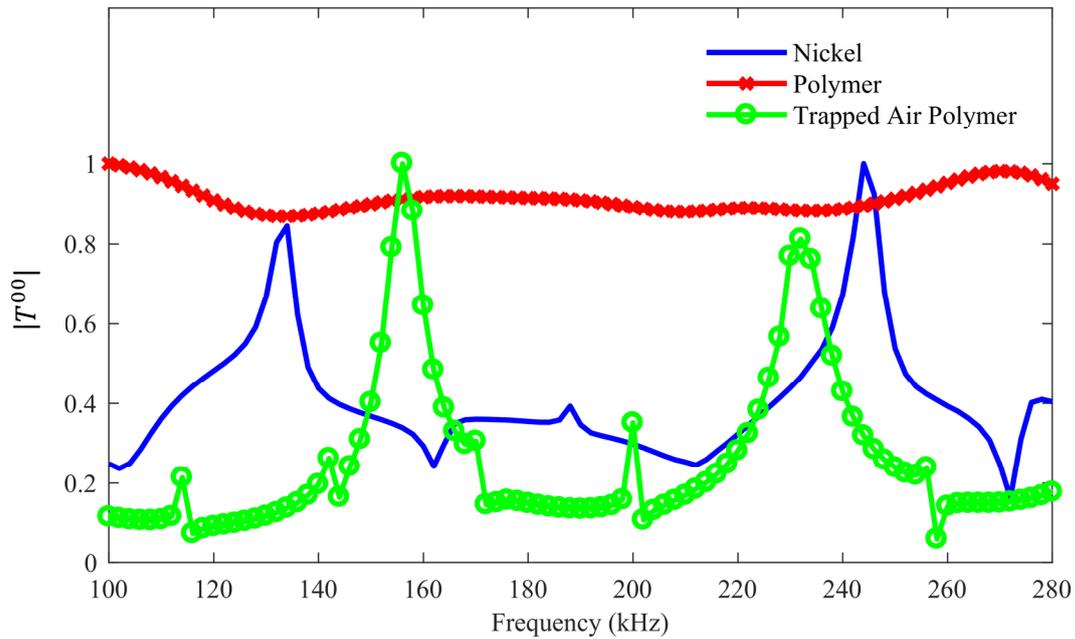

Figure 1: 2-D FEM simulations of $|T^{00}|$ for nickel (blue), polymer (red) and Trapped Air Polymer (green) at different frequencies.

More insight into the impedance mismatch problem is found if $x - z$ sections of the structures are investigated. Fig. 2(a-c) shows the absolute value of the pressure field at peak frequencies for the polymer, nickel and Trapped Air HSAM respectively. While FPRs are well-established and isolated from each other in the trapped-air design, the same it is not true for the polymer substrate, where FPRs are not well-established. This is due to widespread exchange of acoustic energy between individual channels. Finally note that trapped-air polymer designs show more distinct standing waves/FPRs than those of a nickel substrate.

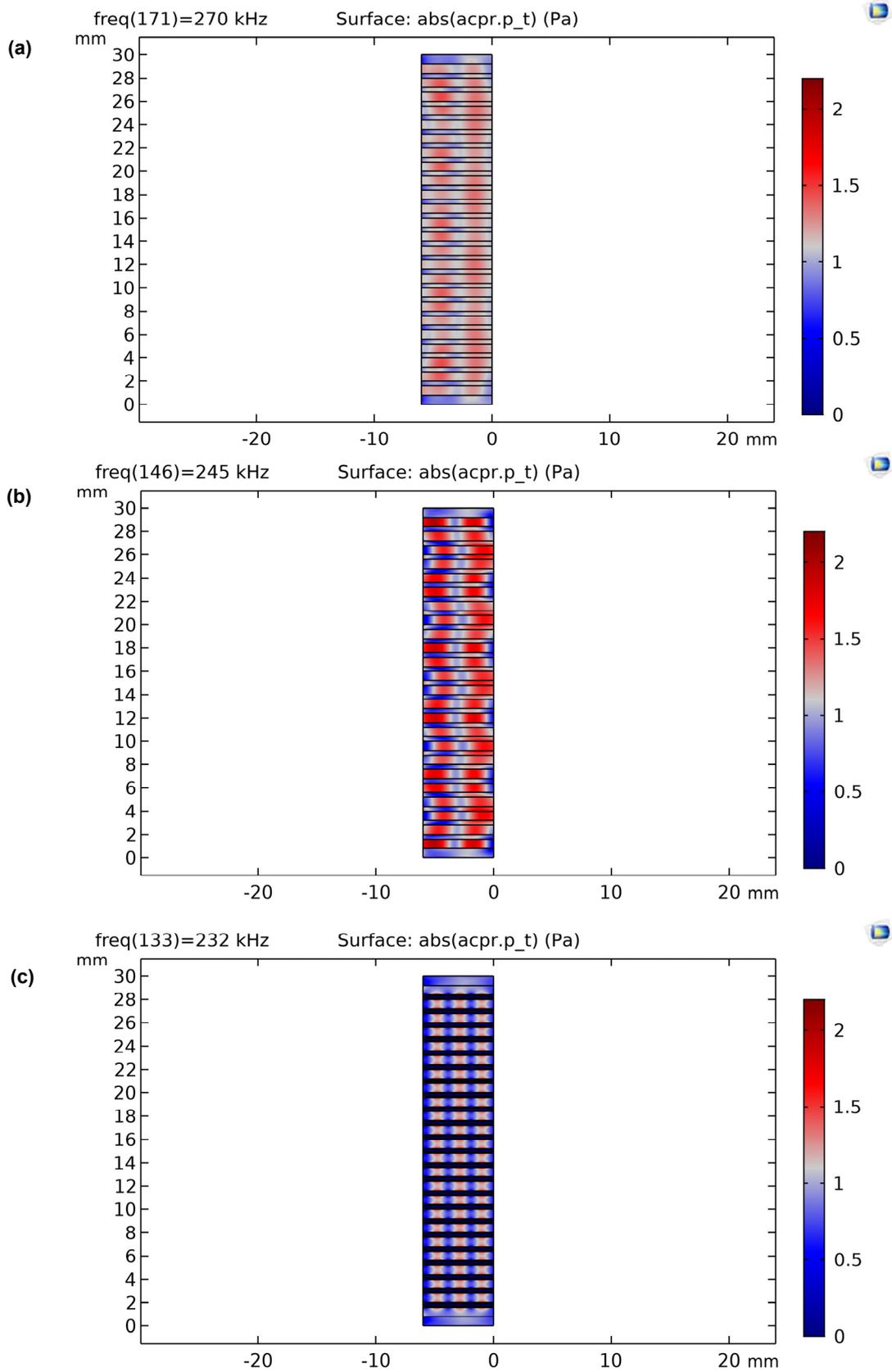

Figure 2: 2-D FEM simulations showing absolute values of the total pressure field for (a) polymer, (b) nickel, and (c) Trapped Air HSAMs designs.

**3-D Finite Element Modelling for Subwavelength imaging**

3-D FEM simulations were conducted to check whether sub-wavelength details could be resolved using the three designs, see Figures 3-4. Figure 3 shows a series of two-dimensional sections of the total pressure field magnitude across $x - y$ planes for the different HSAM designs for a fixed location but at different frequencies. For the polymer HSAM, Figure 3(a), sub-wavelength details are lost due to the global resonance of the whole channeled structure, even enhanced transmission occur around the predicted frequency values. This fact is further appreciated by imaging frequencies closer to the resonance peak, see Fig. 3(b).

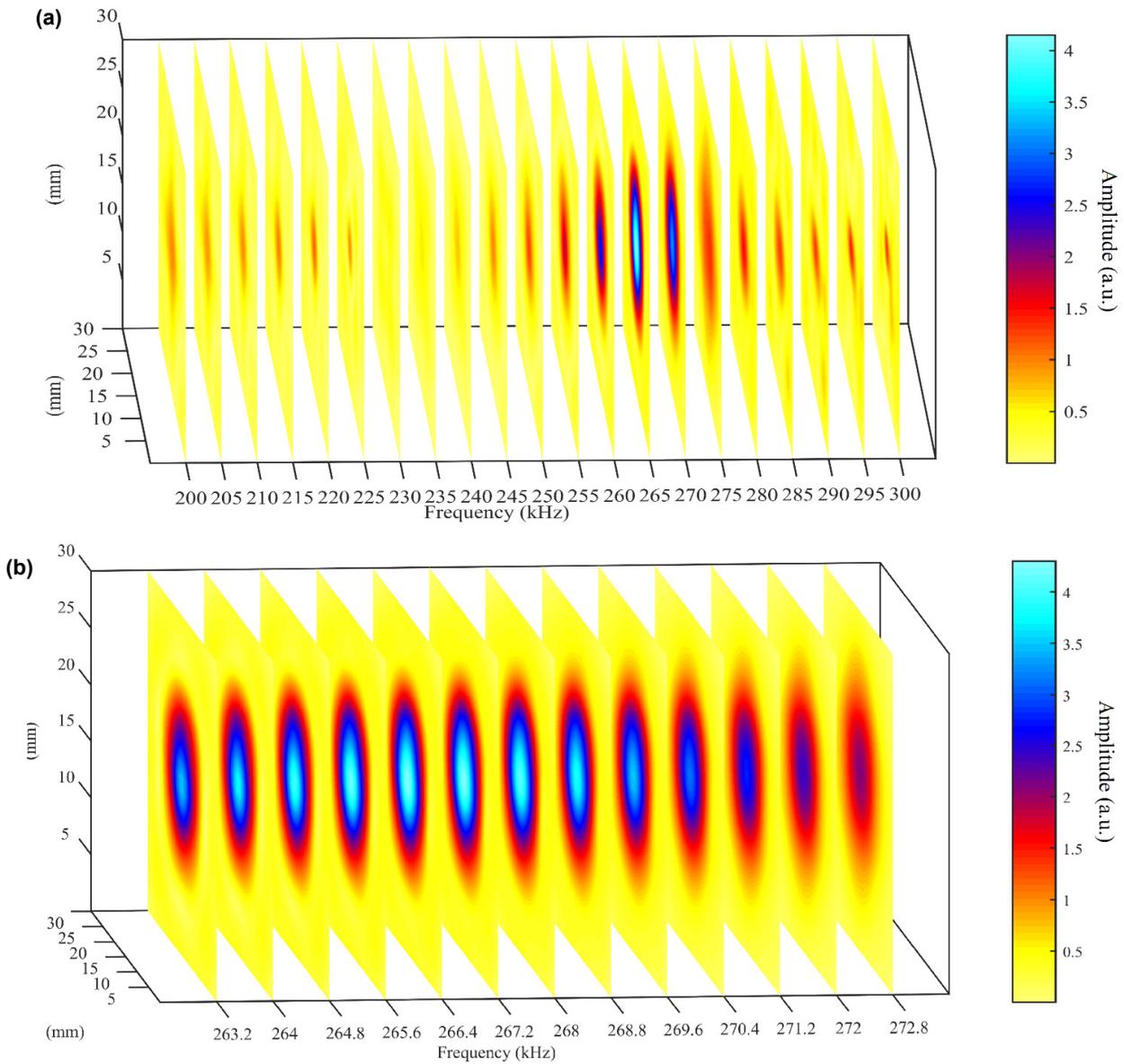

Figure 3: (a) A series of $x - y$ planes at different frequencies for the polymer HSAM obtained from 3-D FEM simulations. "E" details are completely lost at any frequencies in the 200-300 kHz range; (b) details for frequencies around the maximum transmission frequencies values.

Conversely, sub-wavelength imaging capabilities are preserved for both nickel (Fig.4(a)) and Trapped Air HSAMs (Fig.4(b)), the latter showing much better results than the former as all the "E" arms, including the vertical one, are well-imaged.

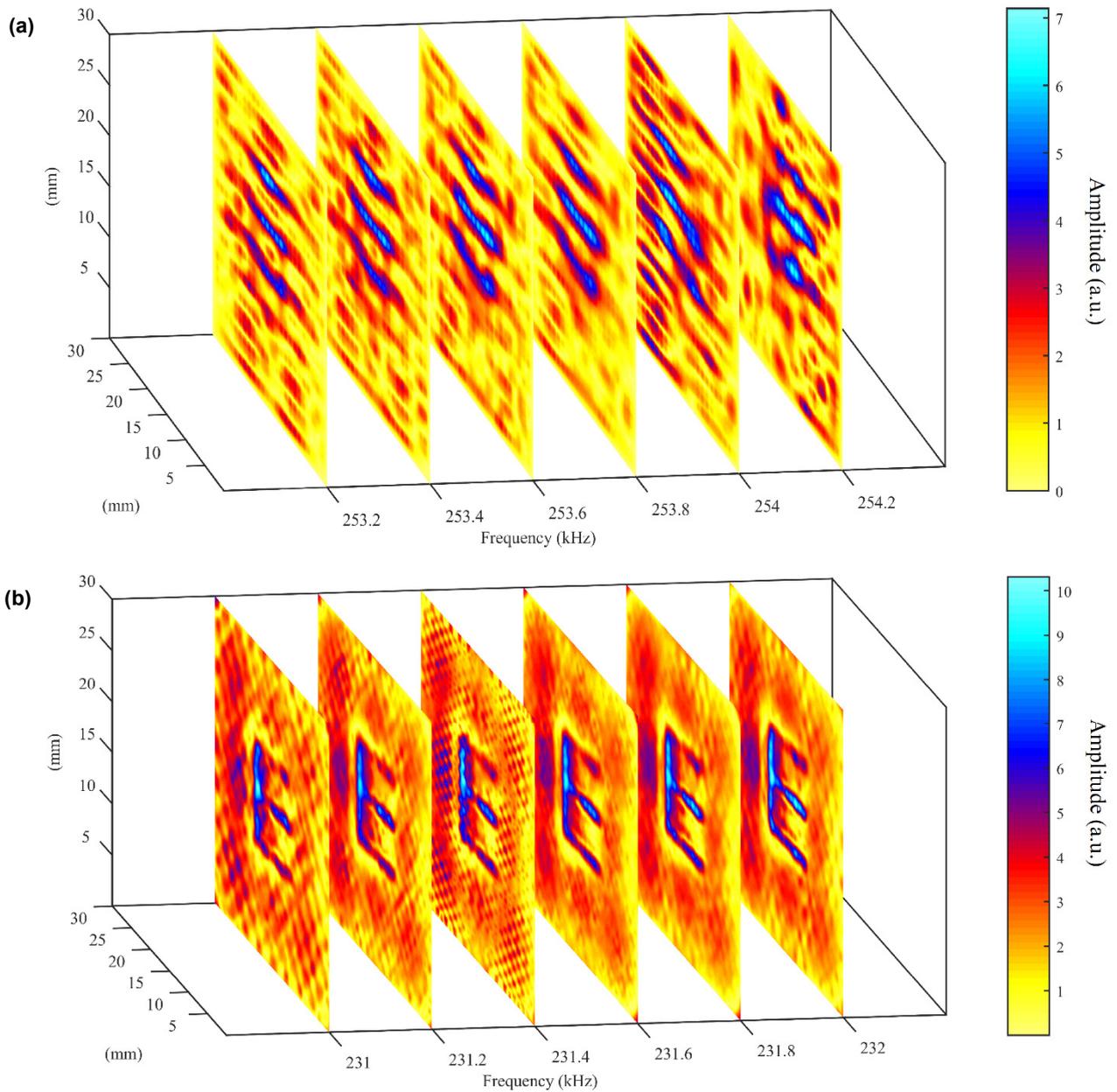

Figure 4: A series of $x - y$ planes at different frequencies for (a) the nickel and (b) the Trapped Air HSAMs for frequencies around the resonance peak obtained from 3-D FEM simulations.

**Experimental Results**

Additional experiments (see Method) were conducted to establish imaging capabilities without the metamaterial, and to compare performance between the new design and metallic- or polymer-based substrates. Figure 5 shows a series of images at different frequencies for the E"-shaped aperture with sub-wavelength dimensions, whereby the hydrophone was placed at 0.1 mm from the slab's outlet without any HSAMs in between. It is noticed that finer details of the "E" are not imaged, as would be expected for a sub-wavelength (1 mm wide) aperture and a detector of finite diameter (0.2 mm).

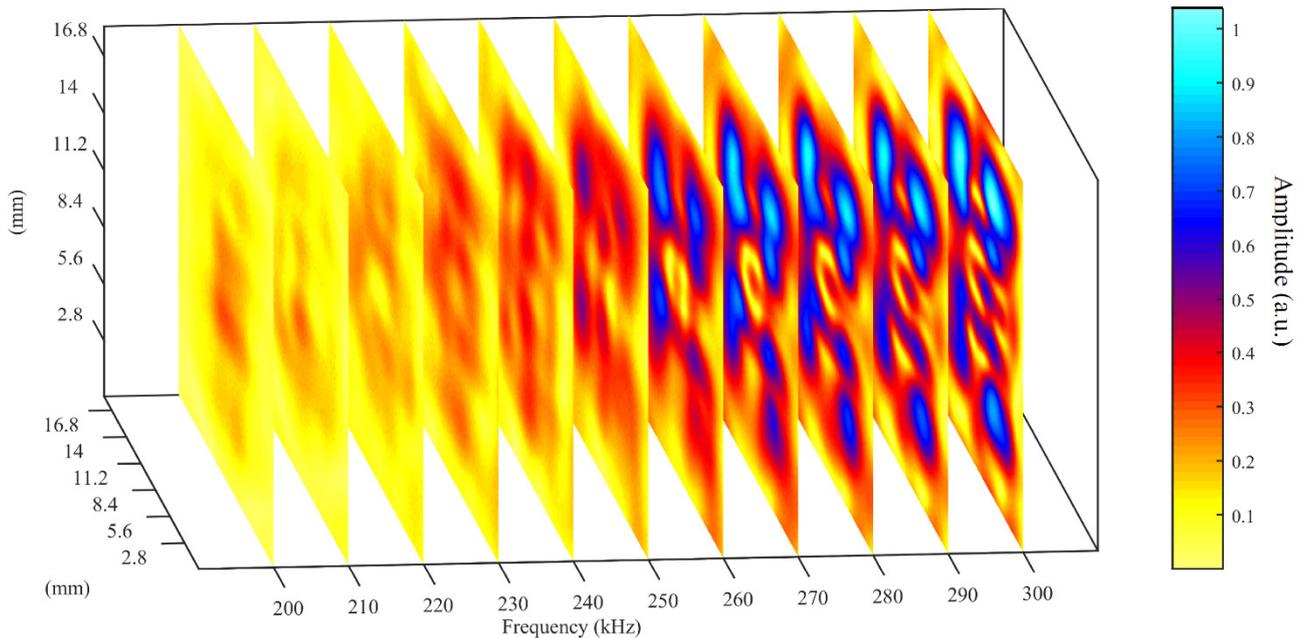

Figure 5: Experimental results for imaging the "E-shaped" subwavelength aperture at a series of $x - y$ planes at different frequencies without a metamaterial being present. The hydrophone was positioned as close as possible to the aperture.

Polymer, nickel, and Trapped Air HSAMs were fabricated by additive manufacturing (see Method) and placed between the "E" aperture and the hydrophone, and photographs are shown in Fig.6 (a-c). Note that there was a 0.1 mm gap between the aperture and the HSAMs.

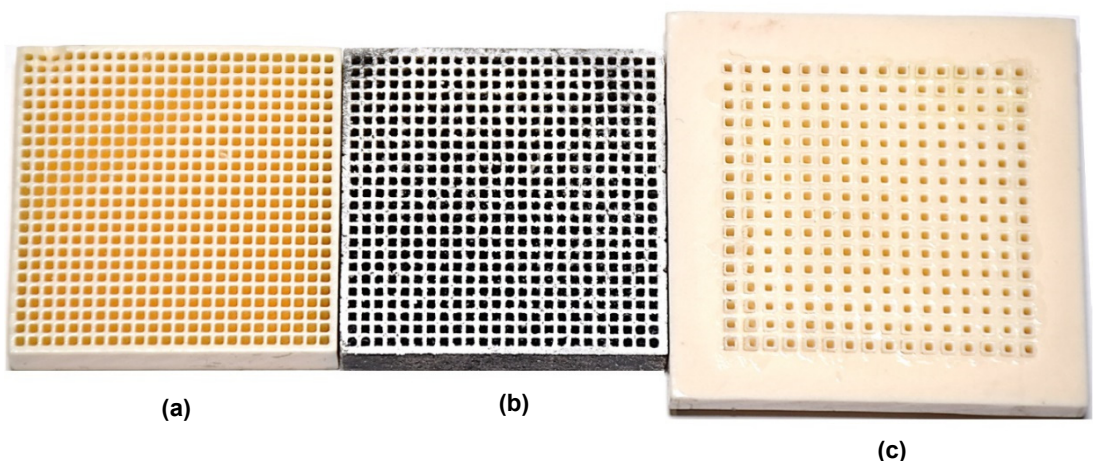

Figure 6: Photographs of the additively-manufactured (a) polymer, (b) nickel, and (c) Trapped Air HSAMs.

Figure 7(a-c) shows the imaging results obtained using the polymer, nickel and Trapped Air HSAM respectively. Notice that "E" details are not imaged by the polymer HSAM, in agreement with the simulations and impedance mismatch prediction (Fig. 7(a)). On the other hand, the finer details of the "E" shaped aperture are seen close to the FPR frequency values for both the nickel and Trapped Air HSAMs. Good results are obtained for the new design, noting that a shift toward the lower frequencies for the best image was obtained experimentally due to its slightly greater thickness ($h \sim$ 6.4 mm).

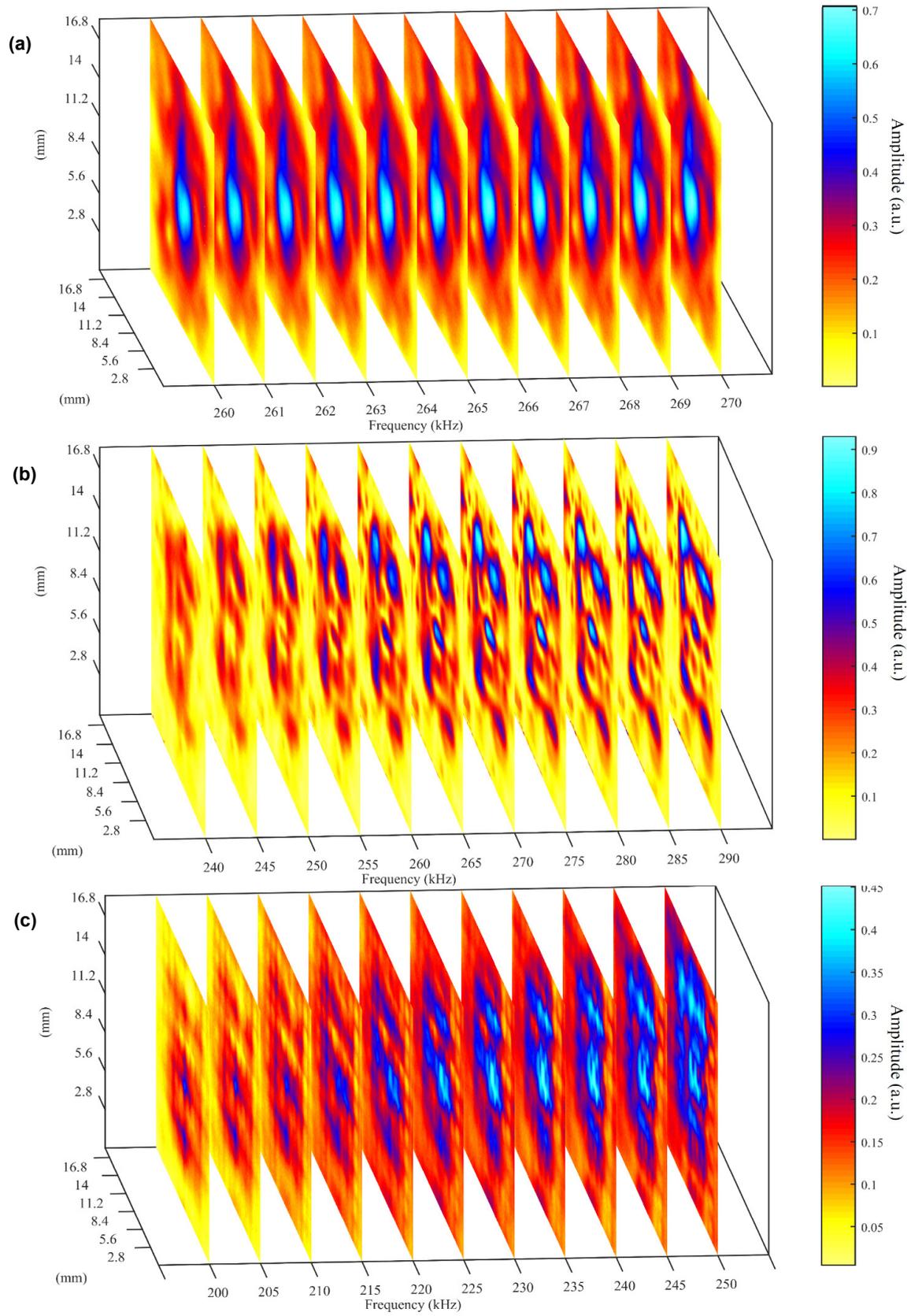

Fig. 7: a series of $x-y$ planes at different frequencies obtained by imaging the "E-shaped" subwavelength aperture with the (a) polymer, (b) nickel, and (c) Trapped Air HSAMs, obtained from experimental data.